\documentclass[12pt]{article}
\usepackage[utf8]{inputenc}
\usepackage{amsmath,graphicx}
\usepackage[style=vancouver]{biblatex}
\usepackage{fullpage}
\usepackage{lineno}
\usepackage{xcolor}
\usepackage{pdflscape}
\usepackage{hyperref}
\usepackage{authblk}

\title{\sc{Efficient inference for differential equation models without numerical solvers}}

\author[1]{Alexander Johnston}
\author[2]{Ruth E. Baker}
\author[1,3]{Matthew J. Simpson}
\affil[1]{School of Mathematical Sciences, Queensland University of Technology (QUT), Brisbane, Australia}
\affil[2]{Wolfson Centre for Mathematical Biology, Mathematical Institute, University of Oxford, Oxford, United Kingdom}
\affil[3]{Centre of Excellence for the Mathematical Analysis of Cellular Systems, QUT, Brisbane, Australia}

\date{}

\begin{document}


\maketitle

\begin{abstract}
Parameter inference is essential when interpreting observational data using mathematical models. Standard inference methods for differential equation models typically rely on obtaining repeated numerical solutions of the differential equation(s). Recent results have explored how numerical truncation error can have major, detrimental, and sometimes hidden impacts on likelihood-based inference by introducing false local maxima into the log-likelihood function. We present a straightforward approach for inference that eliminates the need for solving the underlying differential equations, thereby completely avoiding the impact of truncation error. Open-access Jupyter notebooks, available on \href{https://github.com/AlexanderJohnston1/Generalised-Profiling-Tutorial}{GitHub}, allow others to implement this method for a broad class of widely-used models to interpret biological data.
\end{abstract}

\section{Introduction}

Mechanistic mathematical models, often consisting of systems of differential equations, are essential for understanding various problems in the life sciences. The practical value of developing such models arises when they are used to understand, interpret, and predict observational data. It is therefore essential that mathematical models are related to data via inference. Mathematical models based on ordinary differential equations (ODEs) are fundamentally important for many problems in the life sciences (e.g., biological dynamical systems~\cite{toni2009approximate}, disease dynamics~\cite{wu2023likelihood}, population dynamics~\cite{simpson2022parameter}), and our ability to derive biological insights from these models hinges upon reliable parameter inference in the face of noisy, incomplete data.  Standard inference approaches, both Bayesian and frequentist \cite{wasserman2013all, hines2014determination}, typically rely on repeated numerical solutions of the ODEs, which necessarily introduces truncation error. The introduction of truncation error can have a major, sometimes hidden impact on parameter estimation, since it can introduce false local maxima into the log-likelihood function \cite{creswell2024understanding}. This can occur even when truncation error has a negligible impact on the ODE solution accuracy~\cite{creswell2024understanding}.

We present a straightforward method for inference using ODE models that circumvents the need for solving the ODE model. This approach, called \textit{generalised profiling}~\cite{ramsay2007parameter,ramsay2017dynamic}, avoids issues related to numerical truncation error. It works by using splines to match noisy data, as well as estimating derivatives, allowing us to approximately enforce the governing ODE without solving it. We implement this method for two widely used ODE models and provide easy-to-use, open-access codes on \href{https://github.com/AlexanderJohnston1/Generalised-Profiling-Tutorial}{GitHub} that can be used directly or adapted for other models.

\section{Results and Discussion}

Our approach requires a given set of data to be approximated with splines~\cite{de1978practical}; here we use cubic B-Splines~\cite{BSplineKit}. Consider a set of $IK$ data points, expressed in a vector $y^{\mathrm{o}}(t_{i})$, for $i = 1,2,3, \ldots, IK$, where $K$ types of data are recorded at each of the $I$ time points. We seek to describe these data using an ODE model, for example $\mathrm{d} y/\mathrm{d} t = f(t, y, \theta)$. Here, $y(t)$ is a vector representing the solution of the model, $\theta$ is a vector of unknown parameters that we will estimate, and $f(t,y,\theta)$ is a vector-valued function, with $K$ components, that encodes the biological phenomena of interest (e.g., population dynamics, disease transmission). A distinguishing feature of the procedure is that we never solve the ODE model, meaning that numerical truncation error is not introduced. Moreover, unlike standard inference methods, we do not need to specify initial conditions, since we work with the governing ODE instead of working with the solution. 

To proceed, we construct a spline discretised into $J$ points: $y(t_{j})$, for $j = 1,2,3, \ldots, J$. This will be used to describe the solution of the ODE model, $y(t)$. Choosing $J \gg I$  ensures the spline has a finer resolution than the data. We construct a matrix $B$, of size $J \times IK$, that consists of cubic B-Splines plus a constant function \cite{BSplineKit}. This means that the matrix of basis functions contains a constant vector of all ones. The number of basis functions is chosen to match the number of observations, $IK$. The spline can therefore be written as 
\begin{equation}
\label{B-spline_equation}
    y(t) \approx \sum^{IK}_{i = 1}\beta_{i}b_{i}(t) = B \beta,
\end{equation}
where $b_{i}(t)$ are the basis functions with coefficients $\beta_i$, $B$ is the B-Spline matrix of size $J \times IK$, and $\beta$ is the basis coefficient vector of length $IK$. With this spline representation, we may compute any derivatives of $y(t)$ with respect to the independent variable, $t$, which will be used to approximately enforce the ODE model. 

A critical feature of this methodology is the introduction of two terms in a generalised log-likelihood function that allow us to match the data as well as approximately enforcing the ODE model. To deal with the data, we first make the standard assumption that the noisy data can be interpreted as being drawn from a normal distribution whose mean is the solution of the ODE, $y(t)$, with a constant variance~\cite{murphy2024implementing}. We then introduce the differential operator, $\mathcal{D}$, which allows us to enforce the ODE. For example, if we are dealing with a system of first-order ODEs then $\mathcal{D}(\cdot) = \mathrm{d} (\cdot) / \mathrm{d}t$. To approximately enforce the ODE model, we introduce a regularisation term, giving
\begin{equation}
\label{regularised_likelihood}
	\ell(\theta \mid y^{\mathrm{o}}(t)) = \underbrace{\sum_{i=1}^{IK} \log \left[\phi\left(y^{\mathrm{o}}(t_i); y(t_i), \sigma^2 \right)\right]}_{\rm Data \, Matching} - \underbrace{\vphantom{\sum_{i=1}^{I} \log \left[\phi\left(y^{\mathrm{o}}(t_i); y(t_i), \sigma^2 \right)\right]} w||(\mathcal{D}y - f(t, y, \theta))||_{2}^2}_{\rm Derivative \, Matching},
\end{equation}
where the first term on the right of Eq.~\eqref{regularised_likelihood} provides a measure of how well $y(t)$ matches the data, and the second term on the right of Eq.~\eqref{regularised_likelihood} allows us to approximately enforce the relevant ODE. This approach involves specifying a weight, ${w \geq 0}$, and we will explain how to iteratively determine the weight later. In this expression, $\phi(y; \mu, \sigma^2)$ denotes the probability density function of the normal distribution with mean $\mu$ and variance $\sigma^2$, and $y(t_i)$ is the spline representation of $y(t)$, which we will explain later. The goal of using Eq.~\eqref{regularised_likelihood} is to use maximum likelihood estimation to provide estimates of the model parameters $\theta$, but first we will explain how to approximate $y(t)$ using splines.

To solve for the spline basis functions, we work with a stacked vector, $y^{*}$, of length $(I+J)K$. Elements of $y^{*}$ are $y^{\mathrm{o}}(t_{i})$ for $i = 1,2,3, \dots, IK$, concatenated with $wf(t_{j},y,\theta)$ for $j = 1, 2, 3, \dots, JK$. We also work with a stacked matrix, $B_{\mathrm{obs}}^{*}$, of size $IK \times (I + J)$. This stacked matrix consists of the matrix $B_{\mathrm{obs}}$, of size $IK \times J$, which contains values of $B$, evaluated at each of the observations, $t_{i}$, concatenated with the matrix $w \mathcal{D} B$, of size $IK \times J$, where $\mathcal{D}$ is a $J \times J$ central difference matrix (\href{https://github.com/AlexanderJohnston1/Generalised-Profiling-Tutorial}{GitHub}). Choosing larger values of $w$ places more importance on the degree to which the ODE is enforced relative to the degree to which the data is matched. 

There are two broad steps to the procedure. We first develop initial estimates of $y(t)$ and $\theta$, and then refine these through iteration. The initial ODE solution estimate, $y^{\mathrm{(0)}}(t)$, matches the data without enforcing the ODE model. To achieve this, we set $w^{(0)} = 0$, where the superscript denotes the iteration index. Using Eq.~\eqref{B-spline_equation}, we find an initial solution for the basis function coefficients
\begin{equation}
    \beta^{\mathrm{(0)}}(t) = \left(B_{\mathrm{obs}}^{*, (0)}\right)^{+} y^{*, (0)}(t),
\end{equation}
which can be solved using linear least squares to give an initial estimate for the spline:
\begin{equation}
    y^{\mathrm{(0)}}(t) = B \beta^{\mathrm{(0)}}(t).
\end{equation}
Here $\left(B_{\textrm{obs}}^{*, (0)}\right)^{+}$ denotes the Moore-Penrose inverse of $B_{\textrm{obs}}^{*, (0)}$. This initial estimate has been developed entirely \textit{a posteriori}, and is determined solely by the data. We then provide an initial guess for the model parameters, $\theta^{(0)}$, which allows $f\left(t, y^{(0)}, \theta^{(0)}\right)$ and ${\ell(\theta\mid y^{\mathrm{o}}(t))}$ to be evaluated.

The iterative procedure begins by incorporating the ODE into Eq.~\eqref{regularised_likelihood} using a sufficiently small weight, $w \ll 1$; here taken to be $w^{(1)} = 10^{-2}$. This choice ensures that the iterative procedure commences with a bias towards matching the data. Subsequent iterations produce new estimates of the ODE solution, $y(t)$, and terminate when successive iterates are sufficiently similar. This is achieved by gradually enforcing the ODE model by choosing larger values of $w$, as we now describe. 

We construct a stacked vector, $y^{*, (1)}(t)$, and a corresponding stacked matrix of B-Splines, $B_{\mathrm{obs}}^{*, (1)}$. To construct these, we use $\theta = \hat{\theta}$, the Maximum Likelihood Estimate (MLE), determined using standard numerical optimisation~\cite{NLopt}. This allows the spline representation to be updated. Each subsequent iteration has a corresponding MLE, denoted by $\hat{\theta}^{(n)}$ for the $n$th iteration. Explicitly, for the $n\mathrm{th}$ iteration, where $n\geq1$, we have a stacked vector, $y^{*, (n)}$, of length $(I+J)K$, given by the concatenation of two vectors with components,
\begin{subequations}
\label{nth_augmented_y_vector}
  \begin{align}
    y^{\mathrm{o}}(t_{i}), \hspace{0.4cm} i = 1,2,3, \dots, IK, \quad \textrm{and} \\
    w^{(n)}f\left(t_{j},y^{(n-1)},\hat{\theta}^{(n-1)}\right), \hspace{0.4cm} j = 1, 2, 3, \dots, JK.
  \end{align}
\end{subequations}
Likewise, the stacked matrix $B_{\mathrm{obs}}^{*, (n)}$, of size $IK \times (I + J)$, is given by the concatenation of $B_{\mathrm{obs}}$ and $w^{(n)}\mathcal{D} B$. Using linear least squares, we determine the basis spline coefficients for the $n\mathrm{th}$ iteration
\begin{equation}
    \beta^{(n)}(t) = \left(B_{\mathrm{obs}}^{*, (n)}\right)^{+} y^{*, (n)}(t),
\end{equation}
which yields the $n\mathrm{th}$ spline solution:
\begin{equation}
    y^{(n)}(t) = B \beta^{(n)}(t).
\end{equation}
Successive values of $w^{(n)}$ are chosen by considering the quantity
\begin{equation}
\label{sigma_D_equation}
    {\sigma^2}_{D}^{(n)} = \frac{1}{IK-1} \sum_{i = 1}^{IK} \left[B_{\mathrm{obs}} \beta^{(n)}(t_i) - y^{\mathrm{o}}(t_i)\right]^2,
\end{equation}
which quantifies the extent to which the $n\mathrm{th}$ spline solution matches the data. We also consider the quantity
\begin{equation}
    {\sigma^2}_{M}^{(n)} = \dfrac{1}{JK-1} \sum_{j = 1}^{JK} \left[\mathcal{D} B \beta^{(n)}(t_j) - f\left(t_j,y^{(n)},\hat{\theta}^{(n)}\right)\right]^2,
\end{equation}
which quantifies the extent to which the derivatives of the $n\mathrm{th}$ spline solution match those of the ODE model. Here the subscript $D$ refers to the data while the subscript $M$ refers to the model. Decreasing values of $\sigma_{D}^{(n)}$ indicate that the model matches the data more closely, whereas decreasing values of $\sigma_{M}^{(n)}$ indicates that the spline more closely enforces the ODE.  With this information, we update the weight as follows, 
\begin{equation}
\label{weight_equation}
    w^{(n+1)} = \frac{\sigma_{D}^{(n)}}{\sigma_{M}^{(n)}}, \hspace{0.4cm} n \geq 1.
\end{equation}
This method of calculating $w^{(n+1)}$ ensures that solutions tend towards a compromise between matching the data and enforcing the ODE. 

\subsection*{Case Study I: Harmonic Oscillator}
\begin{figure*}
  \centering  
\includegraphics[width=\textwidth]{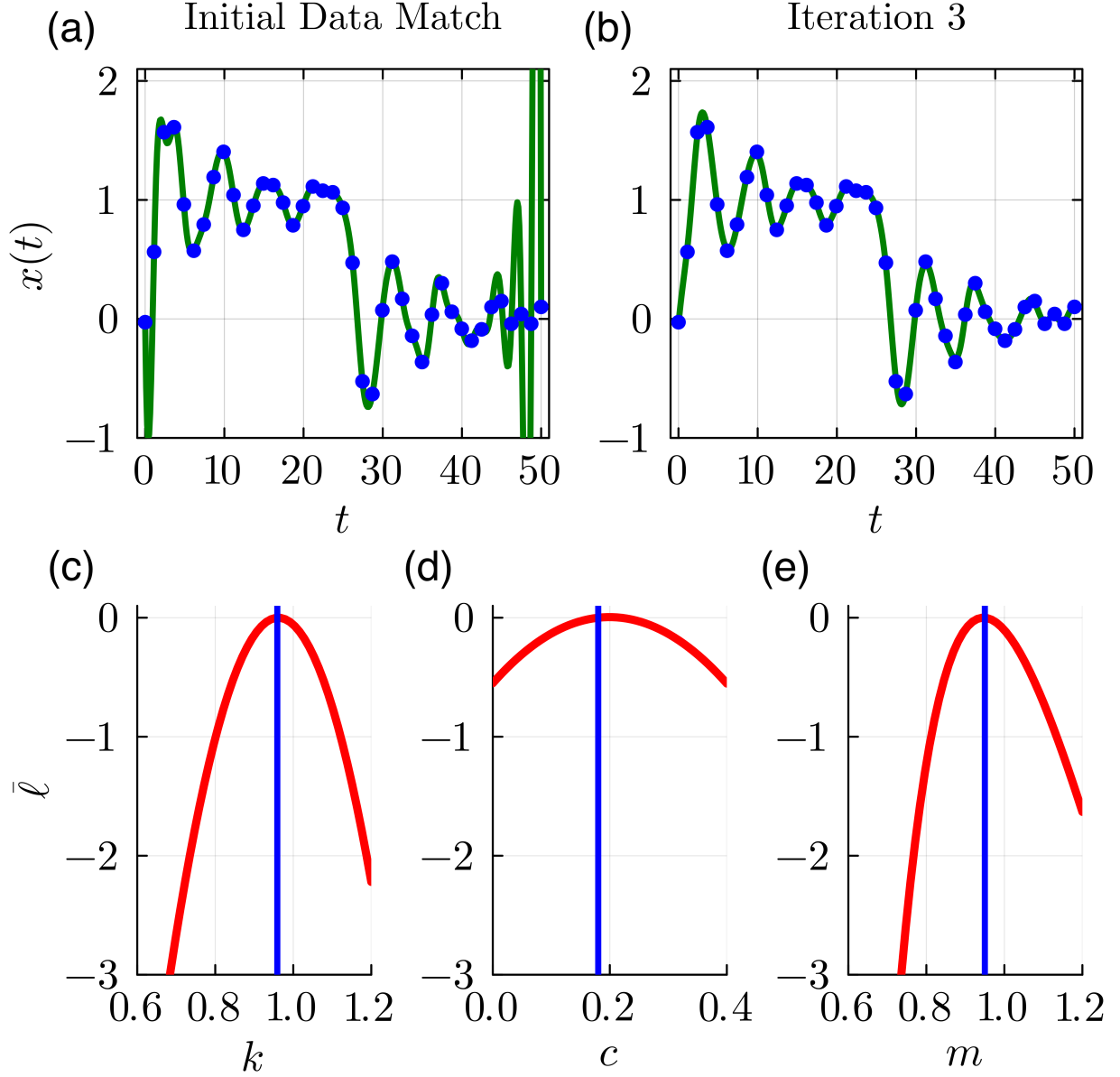}
\caption{Synthetic data (blue dots) and spline solutions (solid green) for the oscillator.  Solutions in (a) and (b) correspond to $n=0$ and $n=3$, respectively. The true parameter values are $(k,c,m) = (1, 0.2, 1)$ and the MLE is $\hat{\theta} = (\hat{k}, \hat{c}, \hat{m}) \approx (0.959, 0.180, 0.950)$. (c)--(e) show slices through the normalised log-likelihood function.  Each slice is superimposed with the corresponding MLE value (solid vertical blue).}
\label{Fig_1}
\end{figure*}

Here we consider parameter inference for the motion of a damped, driven oscillator, 
\begin{equation}
\label{oscillator_equation}
    m \frac{\mathrm{d}^2x}{\mathrm{d}t^2} + c \frac{\mathrm{d}x}{\mathrm{d}t} + kx = F(t),
\end{equation}
where $x(t)$ represents the displacement from equilibrium, and the parameters $\theta = (m, c, k)$ represent the mass, damping coefficient, and spring constant of the oscillator, respectively. Inspired by~\cite{creswell2024understanding}, we specify a discontinuous forcing term, $F(t) = 1 - \mathrm{H}(t - 25)$, where $\mathrm{H}(t)$ is the Heaviside step function. Synthetic data (Supplementary Material) in Fig.~\ref{Fig_1}(a) are superimposed with the initial spline solution, and the same data are superimposed with the spline solution after only three iterations in Fig.~\ref{Fig_1}(b). The initial estimate for $y(t)$ clearly overfits the data as it precisely matches all data points (without any consideration of the governing ODE model). Implementing the procedure for updating weights, described by Eq.~\eqref{weight_equation}, it is visually clear that $y^{(3)}(t)$ ($w^{(3)} \approx 0.357$) provides a good match to the data after only three iterations. The iteration procedure was stopped at this point since we observe that $y^{(3)}(t)$ closely resembles $y^{(2)}(t)$ (\href{https://github.com/AlexanderJohnston1/Generalised-Profiling-Tutorial}{GitHub}).

Results in Figure \ref{Fig_1}(c)-(d) show slices through the normalised log-likelihood function, $\bar{\ell}(\theta | y^{\mathrm{o}}(t))= \ell(\theta | y^{\mathrm{o}}(t)) - \ell(\hat{\theta} | y^{\mathrm{o}}(t))$, where $\ell(\theta | y^{\mathrm{o}}(t))$ is given by Eq.~\eqref{regularised_likelihood}. These results recapitulate and extend those presented by Creswell et al.~\cite{creswell2024understanding}. Each slice is generated by considering the log-likelihood function after three iterations, and then varying one parameter at a time while holding the other two parameters at the MLE. These slices show that the log-likelihood function is smooth, with a clear single maximum value that reasonably approximates the true values, $(k,c,m) = (1,0.2,1)$ (Supplementary Material). We note that these results consider the same model and parameter values considered by Creswell et al.~\cite{creswell2024understanding}, who showed that standard likelihood-based inference for this problem, based on generating repeated numerical solutions of the ODE model, introduces numerical truncation error. In their example, the truncation error led to a non-smooth log-likelihood function characterized by multiple false local maxima, thereby leading to challenges in identifying a unique MLE. 

Unlike standard inference methods, here we estimate both the parameters and the solution, $y(t)$, without ever solving the ODE. A major advantage of taking this approach is that we never have to specify the initial condition(s). For example, to solve  Eq.~\eqref{oscillator_equation} we must specify $x(0)$ and $\mathrm{d}x(0) /\mathrm{d}t$. Standard approaches to inference would either suppose that these quantities are known, fixed values~\cite{creswell2024understanding}, or they would be estimated as part of the inference problem~\cite{simpson2022parameter}. Taking the latter approach would then involve estimating five unknowns, $(m,c,k,x(0),\mathrm{d}x(0) /\mathrm{d}t)$, instead of three. In contrast, we enforce the ODE without solving it, and therefore the initial data can be simply determined from the spline at the final iteration. In the case of the oscillator, our spline gives $x(0) \approx -0.0442$ and $\mathrm{d}x(0) /\mathrm{d}t \approx -0.222$ after three iterations, which are reasonably close to the true values (Supplementary Material). 

\subsection*{Case Study II: Predator-Prey Model}
\begin{figure*}[ht!]
  \centering  
  \includegraphics[width=\textwidth]{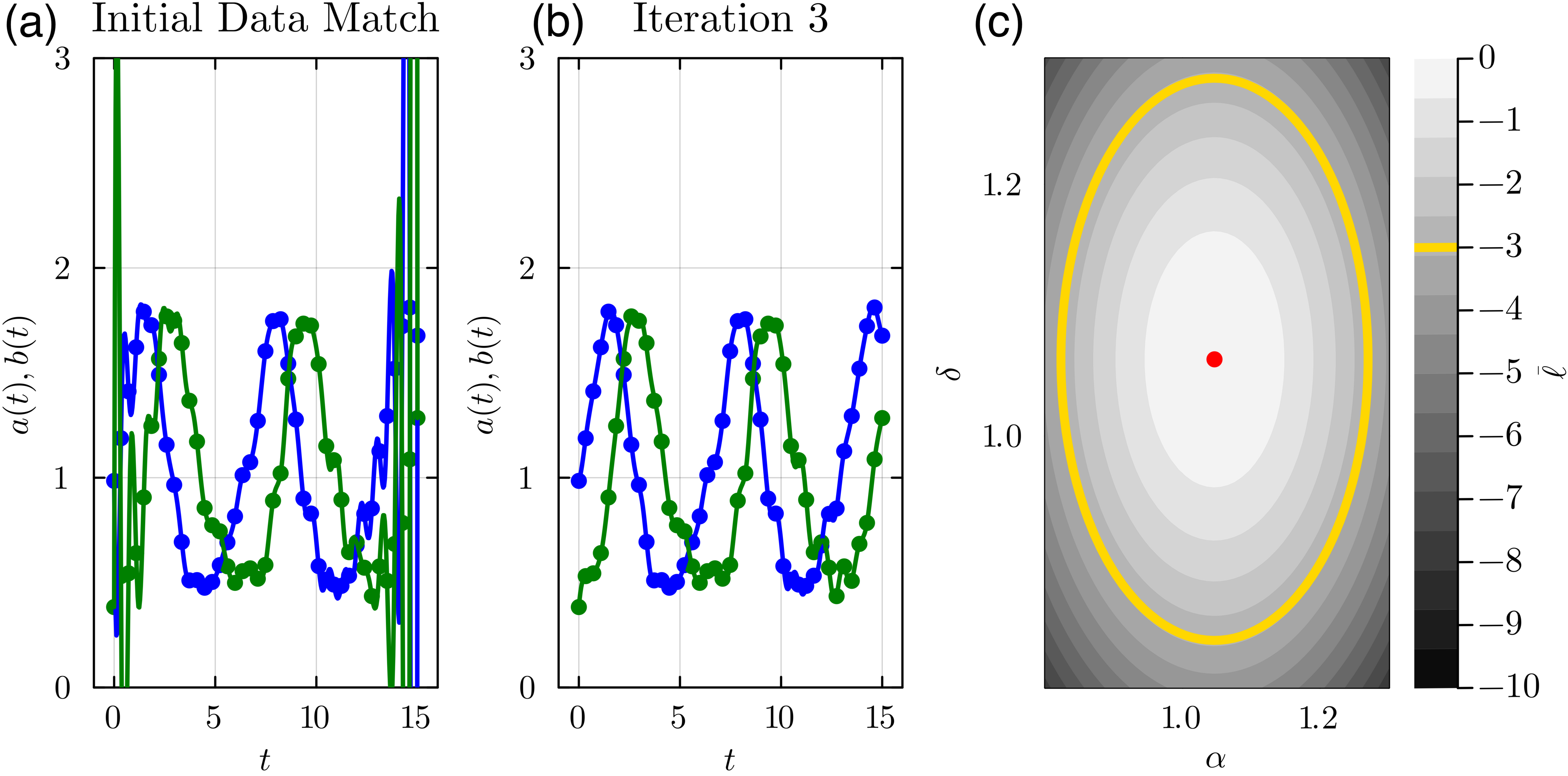}
  \caption{(a),(b): Synthetic data for the predator (green dots) and prey (blue dots) densities superimposed with spline solutions (solid green, solid blue) after $n=0$ and $n=3$ iterations, respectively. (c): Contour plot of the log-likelihood function, $\bar{\ell}(\theta | y^{\rm o}(t))$, superimposed with the MLE (red dot), $\hat{\theta} = (\hat{\alpha}, \hat{\delta}) \approx (1.046, 1.061)$, and the contour defining the 95\% confidence set at $\bar{\ell}^{*} = -\Delta_{0.95,2}/2 \approx  -2.996$ (solid gold).  The true parameter values are $\theta = (\alpha, \delta) =(1,1)$.}
  \label{Fig_2}
\end{figure*}

Here we consider the dimensionless Lotka-Volterra predator-prey model,
\begin{subequations}
\label{Lotka-Volterra_equations}
  \begin{align}
    \frac{\mathrm{d}a}{\mathrm{d}t} &= \alpha a - ab, \\
    \frac{\mathrm{d}b}{\mathrm{d}t} &= \delta ab - b,
  \end{align}
\end{subequations}
where $a(t)$ is the prey density and $b(t)$ is the predator density~\cite{toni2009approximate}, with parameters $\theta = (\alpha, \delta)$. The predator-prey model is nonlinear and involves $K=2$ coupled ODEs. Synthetic data (Supplementary Material) in Fig.~\ref{Fig_2}(a)--(b) is superimposed with splines after $n=0$ and $n=3$ iterations. The solution for $n=0$,  generated with $w^{(0)} = 0$, simply overfits the data without any consideration of the ODE. Selecting weights according to Eq.~\eqref{weight_equation}, we find that after only $n=3$ iterations the splines (generated with $w^{(3)} \approx 0.150$) provide a good visual comparison with the data. In this case we only have two parameters, and we plot $\bar{\ell}(\theta | y^{\mathrm{o}}(t))$ in Fig.~\ref{Fig_2}(c) as a filled contour plot superimposed with the MLE. Furthermore, we show the 95\% confidence set by including the curve  $\bar{\ell}^{*} = -\Delta_{0.95,2}/2 \approx  -2.996$, where the log-likelihood threshold is given by $\bar{\ell}^{*} = -\Delta_{d,q}/2$ where $\Delta_{d,q}$ refers to the $q\mathrm{th}$ quantile of a $\chi^2$ distribution with $d$ degrees of freedom~\cite{royston2007profile}. In this case we see that the MLE, $\hat{\theta} = (\hat{\alpha}, \hat{\delta}) \approx (1.046, 1.061)$, is reasonably close to the true value, $\theta = (\alpha, \delta) = (1,1)$.  Importantly, the log-likelihood function is smooth, and the 95\% confidence region is concentrated locally about the MLE, indicating that the two parameters are well-identified by the data. 

\section{Conclusions and Future Work}

In this report, we present a method of parameter inference for ODE models that does not require the model to be solved (either numerically or analytically). The key element of the approach is to use a regularisation term that allows us to approximately satisfy the governing ODE without explicitly solving the model. Crucially, this approach avoids the impact of numerical truncation errors, which can have severe, detrimental impacts on likelihood-based inference by introducing false local maxima into the log-likelihood function~\cite{creswell2024understanding}. In addition, the method works effectively even when a poor initial guess for the parameters is used. To illustrate the implementation of this method, we apply it to a forced oscillator problem and a simple predator-prey model, and provide open-access software so that others can use the method directly or adapt our code to a wide range of ODE-based inference problems. Future applications of this approach could involve implementing it for: (i) a broader class of differential equation models beyond ODEs; (ii) different kinds of splines for interpolation and derivative estimation; or (iii) different kinds of noise models beyond making the standard assumption that data are normally distributed about the ODE solution~\cite{murphy2024implementing}. A final point is that we take the same approach as Creswell et al.~\cite{creswell2024understanding} by assuming that the variance in the noise model is a known constant. However, our approach also lends itself to estimating the variance~\cite{simpson2022parameter} through Eq.~\ref{sigma_D_equation}.

\section*{Acknowledgments}
The authors thank the mathematical research institute MATRIX in Australia, where some initial work on this project was conducted. The authors also acknowledge Oliver J. Maclaren for providing teaching resources from the University of Auckland, which contain a guide to implement a version of the method presented in the paper.

\section*{Appendix}
This section outlines the process by which synthetic data was generated for the two models in the report. The code that generates these data sets is available on \href{https://github.com/AlexanderJohnston1/Generalised-Profiling-Tutorial}{GitHub}.

\subsection*{Harmonic oscillator}
We consider the model of a damped, driven oscillator given by Eq.~\eqref{oscillator_equation}.

The analytical solution of this model is given by
\begin{equation}
\label{oscillator_solution}
    x(t) = A_0 \mathrm{e}^{-\gamma t} \cos(\phi_0 + \omega t) + \dfrac{F(t)}{k},
\end{equation}
where $\gamma = c/(2m)$, $\omega_0 = \sqrt{k/m}$, $\omega = \sqrt{\omega_0^2 - \gamma^2}$, $\phi_0 = \arctan{\left[(\textrm{d}x(0) /\textrm{d}t)/(\omega(F/k - x(0))) - \gamma/\omega)\right]}$, and ${A_0 = (x(0) - F/k)\cos(\phi_0)}$. We set ${\theta = (k, c, m) = (1,0.2,1)}$ with initial conditions \\ ${(x(0), \textrm{d}x(0) /\textrm{d}t) = (0, 0)}$.
${F(t) = 1 - \textrm{H}(t - 25)}$, where $\textrm{H}(t)$ is the Heaviside step function. 

To generate synthetic data for our study, Eq.~\ref{oscillator_solution} was evaluated at $41$ equidistant data points in the domain $t \in [0, 50]$ and corrupted at each point with independent and identically distributed (IID) Gaussian noise with standard deviation $\sigma = 0.05$. 

\subsection*{Lotka-Volterra equations}
We consider the Lotka-Volterra model of predator-prey dynamics given by Eqs.~\eqref{Lotka-Volterra_equations}.

We generate synthetic data by solving Eqs.~\ref{Lotka-Volterra_equations} using Julia's Tsitouras 5(4) Runge-Kutta method \cite{DifferentialEquations.jl-2017, tsitouras2011runge} with ${\theta = (\alpha, \delta) = (1,1)}$ and with initial conditions $(a(0), b(0)) = (1, 0.5)$. The solution was then evaluated at $41$ equidistant points in the domain $t \in [0,15]$ and corrupted at each point with IID Gaussian noise with $\sigma = 0.05$.

\printbibliography

\end{document}